\RequirePackage[dvipsnames]{xcolor}
\documentclass[journal,twoside,web]{ieeecolor}
\usepackage{tmi}
\usepackage{cite}
\usepackage{amsmath,amssymb,amsfonts}
\usepackage[bb=boondox]{mathalfa}
\usepackage{mathtools}
\usepackage{algorithmic}
\usepackage{graphicx}
\usepackage{stfloats}
\usepackage{textcomp}
\usepackage{arydshln}
\usepackage{booktabs}
\usepackage{hyperref}
\hypersetup{
    colorlinks = true,
    linkcolor = blue,
    anchorcolor = blue,
    citecolor = blue,
    filecolor = blue,
    urlcolor = blue
}

\newcommand\gray[1]{{\color{gray}#1}}
\newcommand\green[1]{{\color{green}}}

\begin{document}
\title{Mixed Supervision of Histopathology Improves Prostate Cancer Classification from MRI }
\author{Abhejit~Rajagopal,
Antonio~C.~Westphalen,
Nathan Velarde,
Tim Ullrich,
Jeffry P.~Simko,
Hao Nguyen,
Thomas~A.~Hope, 
Peder~E.~Z.~Larson,
Kirti~Magudia 
\thanks{A.~Rajagopal, N.~Velarde, T.A.~Hope, and P.E.Z.~Larson are with the Department of Radiology and Biomedical Imaging, University of California, San Francisco, 94158 USA.
J.P.~Simko is with the Department of Pathology, University of California, San Francisco, 94158 USA. 
H.~Nguyen is with the Department of Urology, University of California, San Francisco, 94158 USA.
K.~Magudia is with the Department of Radiology, Duke University, Durham 27705 USA.
T.~Ullrich is with the Department of Diagnostic and Interventional Radiology, University of Dusseldorf, Germany.
A.C.~Westphalen is with the Department of Radiology, University of Washington, Seattle 98105 USA.
Correspondence e-mail: abhejit.rajagopal@ucsf.edu.}}
\markboth{Submitted to IEEE Transactions on Medical Imaging, 2022} 
{Rajagopal \MakeLowercase{\textit{et al.}}: Mixed Supervision of Histopathology Improves Prostate Cancer Classification from MRI }
\maketitle
\begin{abstract}
Non-invasive prostate cancer detection from MRI has the potential to revolutionize patient care by providing early detection of clinically-significant disease (ISUP grade group $\geq 2$), but has thus far shown limited positive predictive value. To address this, we present an MRI-based deep learning method for predicting clinically significant prostate cancer applicable to a patient population with subsequent ground truth biopsy results ranging from benign pathology to ISUP grade group~5. Specifically, we demonstrate that mixed supervision via diverse histopathological ground truth improves classification performance despite the cost of reduced concordance with image-based segmentation. That is, where prior approaches have utilized pathology results as ground truth derived from targeted biopsies and whole-mount prostatectomy to \textit{strongly} supervise the localization of clinically significant cancer, our approach also utilizes \textit{weak} supervision signals extracted from nontargeted systematic biopsies with regional localization to improve overall performance. Our key innovation is performing regression \textit{by distribution} rather than simply by value, enabling use of additional pathology findings traditionally ignored by deep learning strategies. We evaluated our model on a dataset of 973 (testing $n=160$) multi-parametric prostate MRI exams collected at UCSF from 2015-2018 followed by MRI/ultrasound fusion (targeted) biopsy and systematic (nontargeted) biopsy of the prostate gland, demonstrating that deep networks trained with mixed supervision of histopathology can significantly exceed the performance of the Prostate Imaging-Reporting and Data System (PI-RADS) clinical standard for prostate MRI interpretation.
\end{abstract}
\IEEEpeerreviewmaketitle
\section{Introduction}

\IEEEPARstart{A}{lthough} prostate cancer has the highest incidence of any invasive cancer in American men, survival from localized prostate cancer is 100\%~\cite{national2016seer}. Prostate Specific Antigen (PSA) blood tests historically used for prostate cancer screening have high sensitivity but low specificity. Prostate MRI has been widely incorporated into practice in combination with Prostate Imaging-Reporting and Data System (PI-RADS), allowing direct MR-guided biopsy of the prostate gland or more commonly MR/ultrasound fusion biopsy~\cite{rosenkrantz2016prostate,filson2016prostate}. However, PI-RADs demonstrates a wide variability in positive predictive value and overall low positive predictive value for clinically significant prostate cancer (CS-PCa)~\cite{westphalen2020variability}. Thus, there is great interest in developing data-driven deep learning methods to augment prostate MRI interpretation to the benefit of improved detection of CS-PCa. 

\begin{figure*}[ht!]
    \centering
    \includegraphics[width=\linewidth]{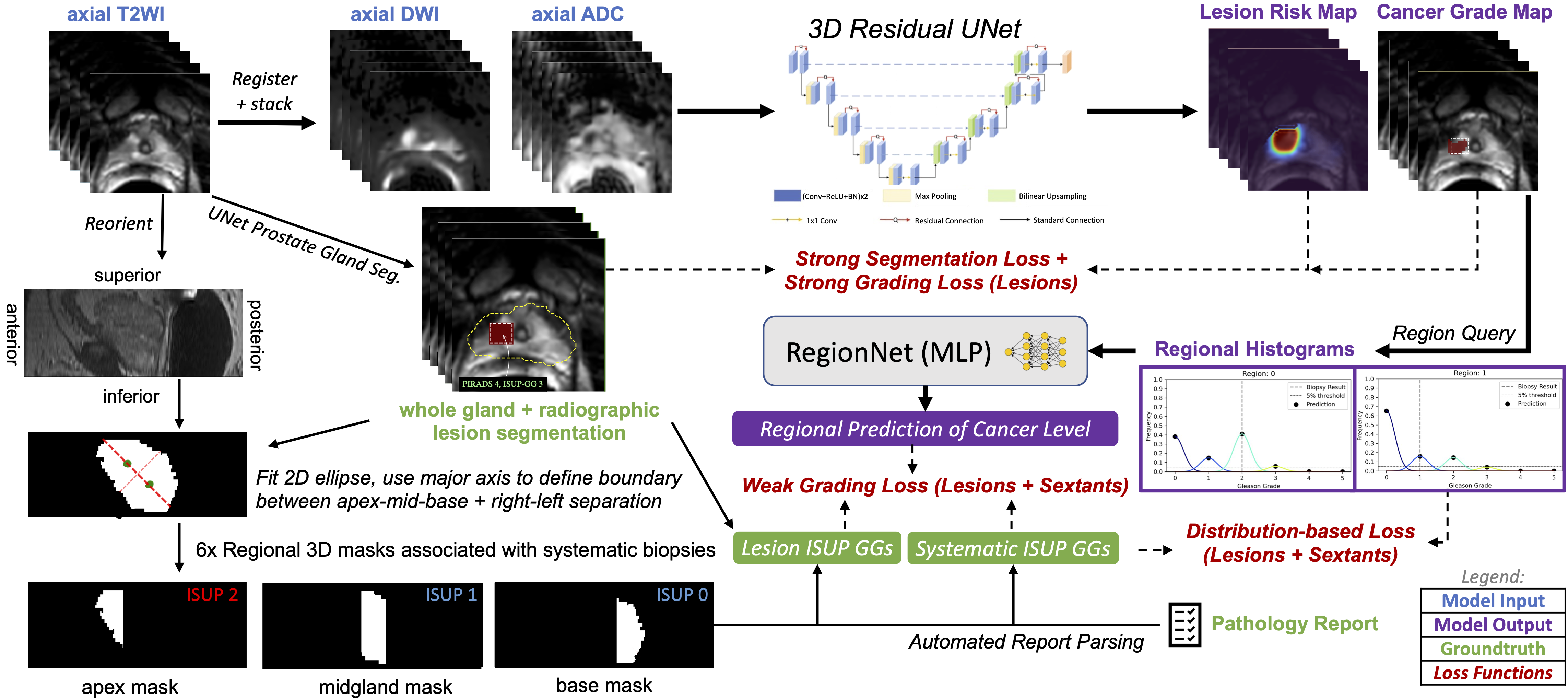}
    \caption{Overview of UCNet model for prediction of prostate cancer from MRI and training via mixed supervision of histopathology ground truth. The 3D model inputs ({\color{MidnightBlue}{blue}}) are fed to a fully-convolutional neural network (here, a 3D Residual UNet) to produce two primary outputs ({\color{RoyalPurple}{purple}}): a Lesion~Risk~Map and a Cancer Grade Map. The primary outputs are compatible with conventional strong voxel-level ground truth for lesions, but incompatible with biopsy data from inhomogeneous tissue in the sextants. To handle this, we query the Cancer Grade Map with geometric masks to produce regional cancer grade histograms, which are weakly supervised using a final RegionNet classification module and a novel distribution-based loss for cancer. 
    }
    \label{fig:overview}
    \vspace{-5mm}
\end{figure*}

Treatment of prostate cancer is driven by biopsying the prostate gland, an invasive procedure with possible complications. Historically, systematic biopsy is used to sample sextants of the prostate gland (left apex, left mid gland, left base, right apex, right mid gland and right base) as estimated by the urologist. MRI/ultrasound fusion biopsy is able to target MRI lesions identified as suspicious on prostate MRI. However, prostate biopsies do not sample the entire gland and may miss small and subtle tumors. Whole-mount analysis of prostatectomy specimens whereby the entire prostate gland is surgically removed is better as the entire prostate gland is analyzed. However, only a small proportion of patients undergo prostatectomy and relying on this population leads to significant sample bias. Ultimately, biopsy data is abundant and better represents the distribution of the population being initially evaluated for prostate cancer by prostate MRI. Thus, an artificial intelligence (AI) strategy with the aim of detecting CS-PCa in the population being screened for prostate cancer must incorporate prostate biospy data. 

Most deep learning strategies focus on MRI-visible disease, which subsequently undergo MRI/ultrasound fusion (i.e.~targeted) biopsies\footnote{However, these prior approaches often inconsistently report results with missing comparisons to PI-RADS or the distribution of tumor grades, leading to ultimately poor generalization performance for developed models~\cite{belue2022low}.}.
In~\cite{schelb2019classification}, authors used lesion biopsy data from and T2-weighted MRI from 312 men with a 2D UNet model, achieving 62-62.5\% balanced accuracy, but found that it was not statistically better than PI-RADS.
In~\cite{cao2019joint}, authors focused on clinically significant cancer in 417 patients receiving prostatectomy, achieving 0.79-0.81 AUC for CS-PCa detection using multi-parametric MRI, but found it was 1.5-3.5\% lower in accuracy than radiologists. Authors in~\cite{de2020deep} used the Prostate-X2 challenge dataset and found a voxel-wise kappa of 0.446$\pm$0.082 using cross-validation, but a lesion-wise kappa of 0.13$\pm$0.27 on the test set. Meanwhile, in~\cite{redekop2021attention} authors showed with 986 exams that image-level labels can achieve 0.75 $\pm$0.03 AUC in grade group classification via attention-based multiple instance learning. Finally, in~\cite{mehralivand2022cascaded} authors used 1043 in-house and 347 Prostate-X exams, and showed that a 3D UNet could predict CS-PCa in prostate MRI lesions at nearly 56.1\% sensitivity and 62.7\% precision, compared to 30.8\% classification accuracy with PI-RADS. 
Despite differences in the reported metric and issues balancing model sensitivity and false-positive rate~\cite{roest2022comparative}, the consensus here is that advanced deep learning techniques can feasibly classify cancer inside suspicious MRI-visible lesions.

Most AI models for prostate cancer detection rely on MRI-visible lesions. However, not all tumors are visible on prostate MRI. For example, a study utilizing our dataset (973 prostate MRI exams conducted at UCSF from 2015-2018 with subsequent MRI/ultrasound fusion biopsy and full systematic biopsy) showed that 14.4\% of cases had a clinically significant upgrade of prostate cancer by nontargeted systematic biopsy compared to MRI targeted biopsy (i.e. MRI visible lesions)~\cite{velarde2022us}. Thus, incorporation of both targeted fusion biopsy for MRI-visible lesions and non-targeted systematic biopsy of disease not necessarily perceived prospectively on MRI, may improve automated algorithms for the detection and risk stratification of prostate cancer.

In this paper, we describe a new approach that can incorporate additional supervision from all biopsy data to DNNs for identifying clinically significant prostate cancer (CS-PCa) from MRI. Specifically, we introduce a unified classification model architecture, 3D UCNet, which is composed of global, regional, and spatial prediction modules for predicting cancer on an exam-wise, region-wise, and voxel-wise basis, respectively~(Fig.~\ref{fig:overview}). Of primary importance in our approach is how we encode and exploit histopathology ground truth. Unlike the cardinal encoding used in~\cite{cao2019joint}, we adopt a multi-class histogram representation that supports regional classification of homogeneous and \textit{inhomogeneous} tissue via a classification head, as well as a novel distribution-based loss function that enables visualization of cancer grade prediction in each region. The handling of inhomogeneous tissue is essential for leveraging systematic biopsy data, since the full extent of the cancerous region sampled by the biopsy core is nonsalient in MRI, and thus the Gleason pattern or ISUP grade group (GG) extracted from pathology report processing must be applied delicately to avoid inconsistency during model training.

\vspace{-2mm}
\subsection{Contributions}
Thus our contributions are as follows:
\begin{itemize}
    \item We introduce 3D UCNet, which features regional and global prediction modules that enable gland-level classification of cancer from a semantic segmentation backbone.
    \item We introduce a histogram-based representation that enables a new type of indirect (weak) supervision consistent with how biopsy cores are annotated by pathologists.
    \item We demonstrate that mixed supervision techniques expand the type of histopathology ground truth used in training and improve CNN-based detection of clinically significant prostate cancer from prostate MRI.
    \item We open source our models, algorithms, and analysis: \url{https://gitlab.com/abhe/prostate-mpmri}
\end{itemize}

\section{Dataset and Assumptions}

\subsection{Multi-parametric Prostate MRI}
Multi-parametric prostate MRI (mp-MRI) is composed of T2-weighted imaging (T2WI), diffusion-weighted imaging (DWI) with an associated apparent diffusion coefficient (ADC) map, and dynamic contrast-enhanced (DCE) MRI. DWI, which typically includes both low and high b-value images, and the associated ADC map represent the restriction of water movement in tissue, which is thought to complement the anatomic T2WI series for detecting prostate cancer~\cite{gupta2020pi}. DCE provides data on the contrast enhancement kinetics of the imaged soft tissues and may help to delineate prostate tumors. All together mp-MRI provides a rich set of information, although  variations due to differences in choice of MR pulse sequence, scanner hardware, and individual patient are commonplace. 

Herein, we used bi-parametric (bp-MRI) consisting of just the T2WI and DWI data~\cite{gatti2019prostate,van2019high}. The T2WI, high B-value DWI and ADC map were registered and resampled uniformly across the dataset to the same spatial resolution of
[1.0, 1.0, 2.24]mm
for x, y, and z axis respectively and normalized. For analysis, these imaging volumes are cropped with centering of the prostate gland and final pixel dimensions of
[64, 64, 32].

\subsection{Histopathology Derived from Prostate Biopsy}
Prostate MRI exams are typically performed after the prostate-specific antigen (PSA) is found to be elevated along with the clinical assessment of a urologist. For our dataset, fellowship-trained abdominal imaging radiologists at UCSF provided routine clinical interpretation of prostate MRI exams with identification of MRI targets, each with an associated PI-RADS scores. These MRI targets were subsequently underwent MRI/ultrasound fusion guided prostate biopsy by a urologist, whereby an axial T2WI is fused by software with live ultrasound to sample MRI targets under ultrasound guidance. In addition to these targeted biopsies, urologists routinely performed systematic biopsies, sampling the prostate in 6 sextants (left apex, left mid gland, left base, right apex, right mid gland, right base). As the exact coordinates of these systematic biopsies are not recorded, these biopsies are only associated with the sextant regions of the prostate gland.  Thus, for our dataset, we assumed the location of the systematic biopsies corresponded to a region defined by the geometric division of the prostate into sextants in the registered MR-image space. For the targeted biopsies, we assumed the location of the targeted biopsy corresponded to a radiologist annotated bounding box around each MRI-identified lesion on each applicable slice~\ref{fig:overview}. The ground truth histopathology for each biopsy sample was determined by a genitourinary pathologist with more than 20 years of experience who graded all specimens using Gleason score~\cite{epstein20162014}. Gleason scores were converted to ISUP grade group (1-5). We used 0 to represent benign pathology results. ISUP grade group $\geq 2$ indicates clinically significant disease.

\subsection{Gland Segmentation and Contrast Normalization}
As T2 and DWI have arbitrary non-quantitative image amplitudes, we applied interquartile range (IQR)-based intra-image normalization to control to the MR image intensity values across research sites and eliminate outlying values created by imaging artifacts. Specifically, each image was normalized to the image-level IQR computed within the 3D prostate gland (annotated by a radiologist or defined by a previously developed neural network segmentation model) according to~\cite{pellicer2022deep}:
\begin{align}
    I_{norm} = \frac{I- \text{percentile}(I, 1)}{\text{percentile}(I, 99) - \text{percentile}(I, 1)}
\end{align}

We subsequently applied Z-score image normalization to each exam to overcome the problem of high variability of intensity distribution between different patients by transforming the intensities to have zero mean and unit variance.

\subsection{Summary of Datasets}
Table~\ref{tab:datasets} summarizes the distribution of ISUP grade groups in the training, validation, and testing cohorts used in this study. Clinical exam data from 2015-2018 was collected at the University of California, San Francisco (UCSF) under IRB approval. As evident, the distribution of ISUP grade group is imbalanced in our dataset.
\begin{table}[hbt!]
    \centering 
    \resizebox{1.0\linewidth}{!} {
    \begin{tabular}{c|c|c|c|}
        \textbf{} & Training & Validation & Testing \\
        \hline
        max ISUP 0   & 92 (13.5\%)  & 17 (17.7\%) & 24 (12.1\%) \\
        max ISUP 1   & 222 (32.7\%) & 27 (28.1\%) & 73 (36.9\%) \\
        max ISUP 2   & 228 (33.6\%) & 30 (31.3\%) & 61 (30.8\%) \\
        max ISUP 3-5 & 137 (20.2\%) & 22 (22.9\%) & 40 (20.2\%) \\
        \hline
        Totals       & 679 (69.8\%) & 96 (9.9\%) & 198 (20.3\%)
    \end{tabular}
    }
    \caption{Number of exams in each ISUP grade group for each cohort.}
    \label{tab:datasets}
\end{table}

All 973 patient exams used in this study included systematic biopsy results from all 6 sextants and at least one targeted-biopsy result.  This results in 7278 biopsy results that were used in training, validation or testing.
Each exam also includes, for each targeted biopsy, a PI-RADS score assigned by a board-certified radiologist when the case was interpreted clinically.

\section{UCNet for Mixed Histopathology Supervision} \label{sec:architecture}
For this problem, we utilized an image-to-image CNN with an additional fully-connected classification output head, a unified classification architecture we call ``UCNet''. Our UCNet implementation takes 3D T2WI, high b-value DWI and ADC map as input and predicts lesion segmentation maps, ISUP grade group maps, and region-wise histograms that are used to determine region-wise and exam-wise cancer severity. We believe this architecture is applicable to many deep learning tasks in medical imaging, but it is particularly well-suited to handling the variety of ground truth histopathology data available for prostate cancer. Of particular importance here is the dynamically-populated multi-task objective we used to train UCNet using highly heterogeneous pathology ground truth collected across the UCSF patient population.

Specifically, we use a fully-convolutional 3D residual UNet \cite{lee2017superhuman} as the backbone for UCNet, which in this case accepts a image tensor $x\in\mathbb{R}^{3 \times X \times Y \times Z}$ as input, and produces voxel-wise tanh-activated lesion maps $\hat{y}_\text{seg}\in\mathbb{R}^{1 \times X \times Y \times Z}$ and softmax-activated grade group membership predictions $\hat{y}_\text{gg}\in\mathbb{R}^{K \times X \times Y \times Z}$, where $K\in\mathbb{Z}_{[0,5]}$ represents the number of unique grade groups chosen for regression. For example, $K=2$ in the case of binary detection of CS-PCa (ISUP grade groups $\geq$ 2). $K$ is an adjustable hyperparameter, but we find $K=2$ yields slightly higher validation accuracy for detection of CS-PCa compared to fully multi-class grading approaches.

\subsection{Voxel-wise Prediction}
We used strong supervision for lesion segmentation and lesion grading, when we had relatively localized information.  

We used a combination of dice and weighted binary cross-entropy losses to supervise the radiological lesion segmentation output, $\hat{y}_\text{seg}$, given the ground truth segmentations $y_\text{seg}$, since this has been shown to achieve higher accuracy~\cite{rajagopal2021understanding}:
\begin{align}
    \mathcal{L}_\text{segmentation}(y_\text{seg}, \hat{y}_\text{seg}) = \mathcal{L}_\text{seg-dice} + \mathcal{L}_\text{seg-BCE} \label{eq:loss_segmentation}
\end{align}
\begin{align}
    \mathcal{L}_\text{seg-dice}(y_\text{seg}, \hat{y}_\text{seg}) = 1 - \frac{y_\text{seg} \cdot \hat{y}_\text{seg}}{y_\text{seg} + \hat{y}_\text{seg} + \epsilon}
\end{align}
\begin{align}
    \mathcal{L}_\text{seg-BCE}(y_\text{seg}, \hat{y}_\text{seg}) =
    \sum_z
    \sum_{x, y_\text{seg}=z}
    \frac{
    \text{BCE}(y_\text{seg}(x), \hat{y}_\text{seg}(x))
    }{|z| \cdot n_z}
\end{align}
where $\epsilon$ is chosen arbitrarily small to prevent overflow, $\text{BCE}$ represents the binary cross-entropy function, $n_z$ represents the number of voxels in class $z\in\{0,1\}$ and $|z|$ is the number of classes. Here, $y_\text{seg}$ is determined from as the voxel-wise max of the region masks $y\in\mathbb{R}^{R \times X \times Y \times Z}$, over regions where the supervision signal in the first column of the histopathology matrix $z\in\mathbb{R}^{R \times 2}$ is 1 to represent MRI-annotated lesions.

Similarly, we applied categorical binary cross entropy loss to each voxel of the predicted grade membership maps, $\hat{y}_\text{gg}\in\mathbb{R}^{K \times X \times Y \times Z}$, given one-hot encoded ground truth histopathology data $y_\text{gg} \in \mathbb{R}^{R \times K}$ from regions $R^+$ where $z$ holds a supervision signal of 1 and a regression grade group that is not NaN. This is an important subtlety that allows supervision of segmentation outputs from patient exams even when the pathological grade group is not known for individual lesions. This strong voxel-wise grading or classification loss is then defined as:
\begin{align}
    \small
    \mathcal{L}_\text{GGmap}(y, z, y_\text{gg}, \hat{y}_\text{gg}) = \frac{1}{|R^+|} &\sum_{r \in R^+}  \sum_{k=1}^K y_\text{gg}[r,k] \log{\hat{y}_\text{gg}[\mathbb{1}_r]_k} \label{eq:loss_strong}
\end{align}
where $\mathbb{1}_r$ represents an indicator (Kronecker delta) function selecting the voxels corresponding spatially to region $r$.

\subsection{Regional Prediction}
Unfortunately, strong voxel-wise supervision signals are not available for all histopathology data types, such as biopsy data. This is because the biopsy core only targets a small portion of the prostate gland, either a lesion or a normal-appearing tissue. For targeted lesion biopsies, we assume that the biopsy sample is representative of a \textit{homogeneous} cancer profile within a tumor that is reasonably localized in the MRI coordinate system, enabling the use of the aforementioned strong classification objective, $\mathcal{L}_\text{strong}$. Systematic biopsies, however, are neither localized nor representative of the cancer profile in \textit{heterogeneous} tissue throughout the gland, preventing the use of such strong voxel-wise objectives.

To this end, we utilized two weak supervision objectives in regions $R^*$ where the supervision signal in $z$ is $\geq 1$:
\subsubsection{Histogram Suppression} \label{subsec:loss_regionhist}
Rather than regressing by value with a voxel-wise classification loss, we regress \textit{by distribution} by computing and penalizing a histogram of predicted voxel-wise grade groups for each region of interest. This can be achieved naturally (maintaining model differentiability) by selecting $K$-dimensional voxels of $\hat{y}_\text{gg}$ using the corresponding region masks in $y$, and computing their $K$-dimensional average, resulting in a set of approximate histograms $\hat{h}\in\mathbb{R}^{R^* \times K}$.

\begin{figure}[hbt!]
    \centering
    \includegraphics[height=2.5in]{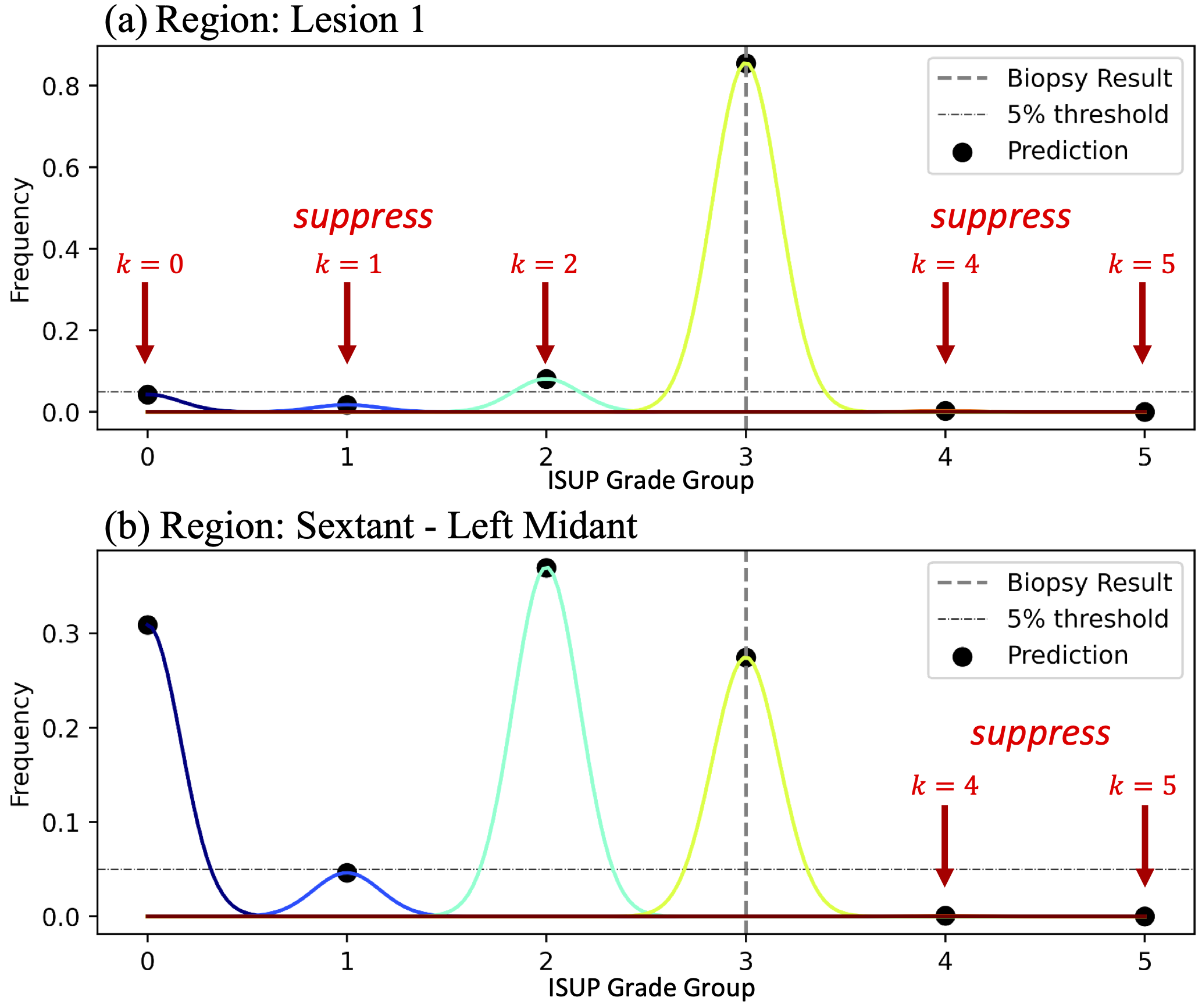}
    \caption{Histogram suppression objectives~(Eq.~\ref{eq:loss_hist_strong}-\ref{eq:loss_hist_high}) penalize predictions of cancer unequal to the ground truth ISUP grade group in homogeneous lesions~(a), and \textit{only above} the histopathological support in inhomogeneous tissue~(b). This form of weak supervision helps shape the predicted \textit{distribution}, particularly in regions where the most frequent prediction may not represent the most clinically-significant cancer grade.}
    \label{fig:histograms}
\end{figure}

For regions $R^\alpha$ where signal $z=1$ (individual MRI target lesions), we assume the biopsy core represents a homogeneous cancer profile, so we penalized non-zero histogram bins that don't correspond to the ground truth grade group~(Fig.~\ref{fig:histograms}a):
\begin{align}
    \mathcal{L}_\text{hist-strong}(z, y_\text{gg}, \hat{h}) = \frac{1}{|R^\alpha|} &\sum_{r \in R^\alpha}  \sum_{k=1}^K y_\text{gg}[r,k] \log{\hat{h}[r,k]} \label{eq:loss_hist_strong}
\end{align}

For regions $R^\beta$ where signal $z>1$ (sextants, where only the highest grade is known), we assume the biopsy core represents the highest cancer in a heterogeneous tissue, so we only penalized non-zero histogram bins that represented grade groups \textit{higher} than the ground truth regression grade group~(Fig.~\ref{fig:histograms}b):
\begin{align}
    \mathcal{L}_\text{hist-high}(z, y_\text{gg}, \hat{h}) = \frac{1}{|R^\beta|} &\sum_{r \in R^\beta}  \quad \sum_{
        \mathclap{\substack{k \; > \\
        \underset{k}{\mathrm{argmax}} \; y_\text{gg}[r] 
        }}
    }^K
    y_\text{gg}[r,k] \log{\hat{h}[r,k]} \label{eq:loss_hist_high}
\end{align}

The net effect of these losses $\mathcal{L}_\text{GGmap-hist} = \mathcal{L}_\text{hist-strong} + \mathcal{L}_\text{hist-high}$ is to suppress the proportion of voxels representing ISUP grade groups not supported by the histopathology data and its expected uncertainty. Note that when $\mathcal{L}_\text{hist-high}=0$, the  histogram bin corresponding to the ground truth grade group may not necessarily have the highest frequency in the region, in agreement with the expectation for heterogeneous tissue.

\subsubsection{Regional Grading} \label{subsec:loss_regionclass}
To provide a clear indication of ISUP grade group in such cases, without relying on manual interpretation of the histograms, we fed the regional histograms $\hat{h}$ into a small dense network with ReLU activation in the final layer. This results in multi-hot encoded vectors $\hat{z}^{R^* \times K}$, indicating the presence of various ISUP grade groups in regions $R^*$.

To maintain the clearest interpretation of $\hat{h}$, we used a single-layer ReLU-activated network with tunable bias and weight-matrix frozen as identity $I_K$. The bias, therein representing a set of ``optimal'' thresholds for declaring clinically-significant prostate cancer in a given region, is optimized through backpropagation, just as other parameters in the fully-convolutional portion of UCNet, via the loss:

\vspace{-4mm}
\begin{small}
    \begin{align}
        \mathcal{L}_\text{region-classifier}(z, y_\text{gg}, \hat{z}) = \frac{1}{|R^*|} &\sum_{r \in R^*}  \sum_{k=1}^K y_\text{gg}[r,k] \log{\hat{z}[r,k]} \label{eq:loss_region_class}
    \end{align}
\end{small}


\subsection{Multi-task Objective}
The overall multi-task objective is:
\begin{align}
    \mathcal{L}(x, y, y_\text{gg}, z) = 
    \alpha_1 \lambda_1 \mathcal{L}_\text{region-classifier}
    + \alpha_2 \lambda_2 \mathcal{L}_\text{GGmap-hist} \label{eq:objective} \\
    + \alpha_3 \lambda_3 \mathcal{L}_\text{GGmap}
    + \alpha_4 \lambda_4 \mathcal{L}_\text{segmentation}  \nonumber
\end{align}
where we empirically choose $\bar{\lambda}=[1.0, 0.5, 1.0, 1.0]$, and $\bar{\alpha}$ is chosen based on the availability of different annotation ground truth available for each exam in a minibatch, as well as the type of supervision signals desired in the experiment. Herein, we choose $\bar{\alpha}=[1.0, 1.0, 1.0, 1.0]$ to use all described objectives whenever possible. For training, we used a batch size of 6 MRI exams and the AdamW optimizer with initial learning rate of $0.0001$, determined empirically by repeatedly sweeping the initial learning rate.

\subsubsection{Objective-Aware Balanced-Batching}
To achieve the best performance, we curated the distribution of each batch by round-robin sampling of MRI exams where the highest ISUP grade group for any region is 0 (negative), 1 (benign), 2 (clinically-significant), 3-5. We find that this empirically leads to better efficiency in training the multi-task objective, since there are more representative samples per task per iteration.

\subsubsection{Global Prediction}
The UCNet architecture supports global prediction modules, which can be based either on the fully-encoded features produced by the encoding branch or the spatial map that is rendered by the decoding branch of the fully-convolutional backbone. We selected the later to focus on improving lesion-wise segmentation and classification accuracy, and to compute a global prediction by maximum filtering of the individual lesion predictions.

\subsection{Inference}
Inference is performed in a similar fashion to training, except with all augmentation turned off. Note that, due to the nature of weak histopathology ground truth, it is not feasible to train UCNet using volumetric patches. Instead, we feed full (registered and resampled) multi-contrast prostate MRI exams as input to UCNet, producing a set of predictions corresponding to all volumetric regions of the prostate gland.

\section{Experiments} \label{sec:experiments}
To determine the potential advantage of mixed-supervision deep learning methods for detection of clinically significant prostate cancer, we designed experiments that represent variations in the supervision methods used during training, as well as alternative classification strategies used during inference or evaluation. Specifically, we choose to withhold or include various types of annotation ground truth, including lesion segmentation masks, voxel-wise ISUP grade group in lesions derived from MR-targeted biopsy and expert histopathological analysis, and maximum ISUP grade group in regions derived from systematic biopsy and expert histopathological analysis. For the regions with biopsy ground truth, in particular, we additionally compare two types of supervision: weak classification-based supervision~(Section~\ref{subsec:loss_regionclass}) and a novel weak \textit{distribution}-based supervision~(Section~\ref{subsec:loss_regionhist}).

Each experiment can be identified by a unique experiment ID, represented by a 4-bit vector, corresponding exactly to the $\bar{\alpha}=\{ \alpha_1, \alpha_2, \alpha_3, \alpha_4 \}$ in Equation~\ref{eq:objective}. However, there are numerous other differences implied for inference when changing supervision methods, as elucidated in the following.

\subsection{Experiment 0001 - lesion segmentation only}
$\alpha_4=1$, $\alpha_{1,2,3}=0$ represents a baseline lesion segmentation experiment, where the combined class-balanced BCE-loss and dice loss (Eq.~\ref{eq:loss_segmentation}) is the sole loss term in the objective. The network model employed is simply a 3D residual UNet, representing the backbone of the UCNet architecture. 

\subsection{Experiment 0011 - lesion segmentation and classification}
$\alpha_{3,4}=1$, $\alpha_{1,2}=0$ represents a baseline lesion segmentation \textit{and} classification experiment, utilizing the same 3D residual UNet backbone of UCNet. The loss terms in the training objective are the segmentation loss (Eq.~\ref{eq:loss_segmentation}) as well as the strong voxel-wise cross entropy objective (Eq.~\ref{eq:loss_strong}). Here, classification decisions are based on the most frequent or mean value in each region based on thresholded voxels of $\hat{y}_\text{GGmap}$. Note, for binary classification the mean and mode inference methods are equivalent. This experiment serves as an important baseline for subsequent methods that utilize the same UNet backbone, but with additional supervision signals.

\subsection{Experiment 0111 - lesion segmentation, lesion classification, and regional classification with histogram suppression}
$\alpha_{2,3,4}=1$, $\alpha_{1}=0$ represents a mixed-supervision objective, novel for its use of distribution-based representations of histopathology. In this experiment, the prediction $\hat{y}_\text{GGmap}$ is penalized in regions where the \textit{maximum} ISUP grade is higher than the ground truth (Eq.~\ref{eq:loss_hist_high}), e.g.~when only the maximum is known for biopsy cores in sextants that represent heterogeneous tissue (i.e.~a mixture of background prostate tissue with clinically significant prostate cancer if it were present). When the biopsy is derived from apparent homogeneous tissue, such as MRI-identified lesions, voxels are penalized in histogram bins both \textit{higher} and \textit{lower} (Eq.~\ref{eq:loss_hist_strong}) than the ground truth maximum. Specifically, the addition of Equations~\ref{eq:loss_hist_strong}-\ref{eq:loss_hist_high} encourages regression \textit{by distribution} rather than regression by value, representing the first (to our knowledge) deep learning method to incorporate the distribution of cancer histopathology. Note that the number of free model parameters is equal in Experiment 0001, 0011, and 0111.
In this setting, and to maintain a clear comparison to the strongly-supervised case, classification is made in the same fashion as in Experiment~0011 by computing the mean or mode value in each region based on thresholded voxels of $\hat{y}_\text{GGmap}$.

\subsection{Experiment 1111 - lesion segmentation, lesion classification, regional classification with histogram suppression, and regional classification with RegionNet}
$\alpha_{1,2,3,4}=1$ represents another mixed-supervision objective that combines the histogram-based penalization of Experiment 0111 together with traditional weak instance-wise or region-wise classification supervision~(Eq.~\ref{eq:loss_region_class}). A region-wise classification is produced by passing the histogram representation of each region to a small neural network, RegionNet. To drive the clearest interpretation and comparison to Experiment 0111, we choose RegionNet as a dense ReLU-activated 1-layer neural network of input and output dimension $K$ and frozen with weight matrix $I_K$, so the $K$ free bias terms represent ``optimal'' thresholds for declaring clinically-significant cancer in each region. Classification in this setting is achieved by finding the most-significant bit (MSB) of the $K$-bit output $\lceil{\hat{y}_\text{gg,r}}\rceil \in \mathbb{Z}^K$ for each region $r$, which represents the highest level of cancer. 

This experiment aims to show that accounting for tissue heterogeneity can dramatically improve classification accuracy of deep learning cancer classification models. Specifically, the combination of distribution-based penalization (Experiment~0111), together with optimal thresholding and MSB-based classification is designed to enable detection of cancer in heterogeneous tissue where the highest level of cancer is \textit{not} the most prevalent (e.g.~mean or mode). 

\begin{table*}[b]
    \centering
    \caption{Comparison of Multi-Task Objectives on withheld test set using as well as PIRADs predictions. Overall class-balanced accuracy, with specificity and sensitivity in brackets.  Experiment IDs represent varying supervision: 0001 = lesion segmentation only; 0011 = lesion segmentation and classification; 0111 = lesion segmentation, lesion classification, and regional classification with histogram suppression; 1111 = lesion segmentation, lesion classification, regional classification with histogram suppression, and regional classification with RegionNet.} \label{tab:results}
    \resizebox{0.95\linewidth}{!} {
\begin{tabular}{rcccc:cc}
\toprule
Experiment ID &    0001 &    0011 &    0111 &    1111 & PI-RADS~$\geq4$ & PI-RADS~$\geq5$ \\
\midrule
Lesion Segmentation IoU                   & 0.220 & 0.188 & 0.162 & 0.150 & \gray{-- --} \\
Lesion Accuracy                           & \gray{-- --} & 59.5\% [94.0\%, 25.0\%] & 61.1\% [93.0\%, 29.2\%] & \textbf{70.3\%} [76.6\%, \textbf{63.9\%}] & 62.4\% [31.8\%, 93.1\%] & 66.7\% [82.5\%, 52.4\%] \\
Gland Accuracy (via Lesions)         & \gray{-- --} & 55.8\% [90.7\%, 20.8\%] & 57.7\% [90.7\%, 24.8\%] & \textbf{62.8\%} [70.1\%, \textbf{55.4\%}] & 61.0\% [30.9\%, 91.1\%] & 64.4\% [79.4\%, 49.5\%] \\
\bottomrule
\end{tabular}
}
\end{table*}

\subsection{Training Details and Stopping Criteria}
For each experiment, UCNet was trained twice: (1)~by performing an initial learning rate sweep and selecting an optimal point based on decrease in the objective, and (2)~by fixing the initial learning rate to 0.0001. Training was performed for 500 epochs, and the checkpoint with the maximum lesion-wise \textit{validation} accuracy was chosen. In all cases, the fixed learning rate run achieved superior validation performance, and these models were thus chosen for test-set evaluation and reporting.

\subsection{Performance Metrics}

\subsubsection{Lesion Segmentation}
Lesion segmentation accuracy is measured by the standard intersection-over-union (IoU) metric, measuring the overlap between the radiologist annotation (performed as a bounding box) and the predicted binarized lesion segmentation map $\hat{y}_\text{seg}$. Although IoU tends to underestimate the lesion segmentation performance, we believe this is provides an accurate measure of model confidence in segmentation on this challenging task. Note that the assumed radiologist ground truth is not based on pathology, so the IoU metric is only serving to measure the overlap or consistency between the model and the radiologist.  Furthermore, the bounding box annotations are imprecise which will also reduce the IoU performance, especially for small non-spherical lesions.

\subsubsection{Lesion Classification Accuracy}
Unlike many prior works that measure area under the receiver-operating characteristic curve (AUC) via cross-validation~\cite{schelb2019classification,redekop2021attention,roest2022comparative}, here we measure overall \textit{class-balanced} lesion classification accuracy on a withheld test set. Class-balanced accuracy is computed simply as the arithmetic mean of the true-positive rate~(TPR) and true-negative rate~(TNR), or in the multi-class setting, as the arithmetic mean of the trace of confusion matrix. 
We use class-balanced accuracy for two reasons: (1)~we believe this is a more representative measure of model performance because it weights positive and negative cases equally, and (2)~it captures model performance without any post-training tuning of the final threshold. For clarity, we note that the TPR and TNR represent sensitivity and specificity, respectively.

\subsubsection{Gland Classification Accuracy}
Gland accuracy is measured in a similar fashion to lesion classification accuracy using class-balanced accuracy metrics. Specifically in the binary CS-PCa setting, the maximum prediction for all lesions (or regions, as appropriate) in an exam is compared to the ground truth maximum to yield a true-positive or true-negative count. The TPR and TNR are subsequently averaged to yield the overall class-balanced gland classification accuracy.

\section{Results}
\begin{figure*}[b]
    \includegraphics[width=\linewidth]{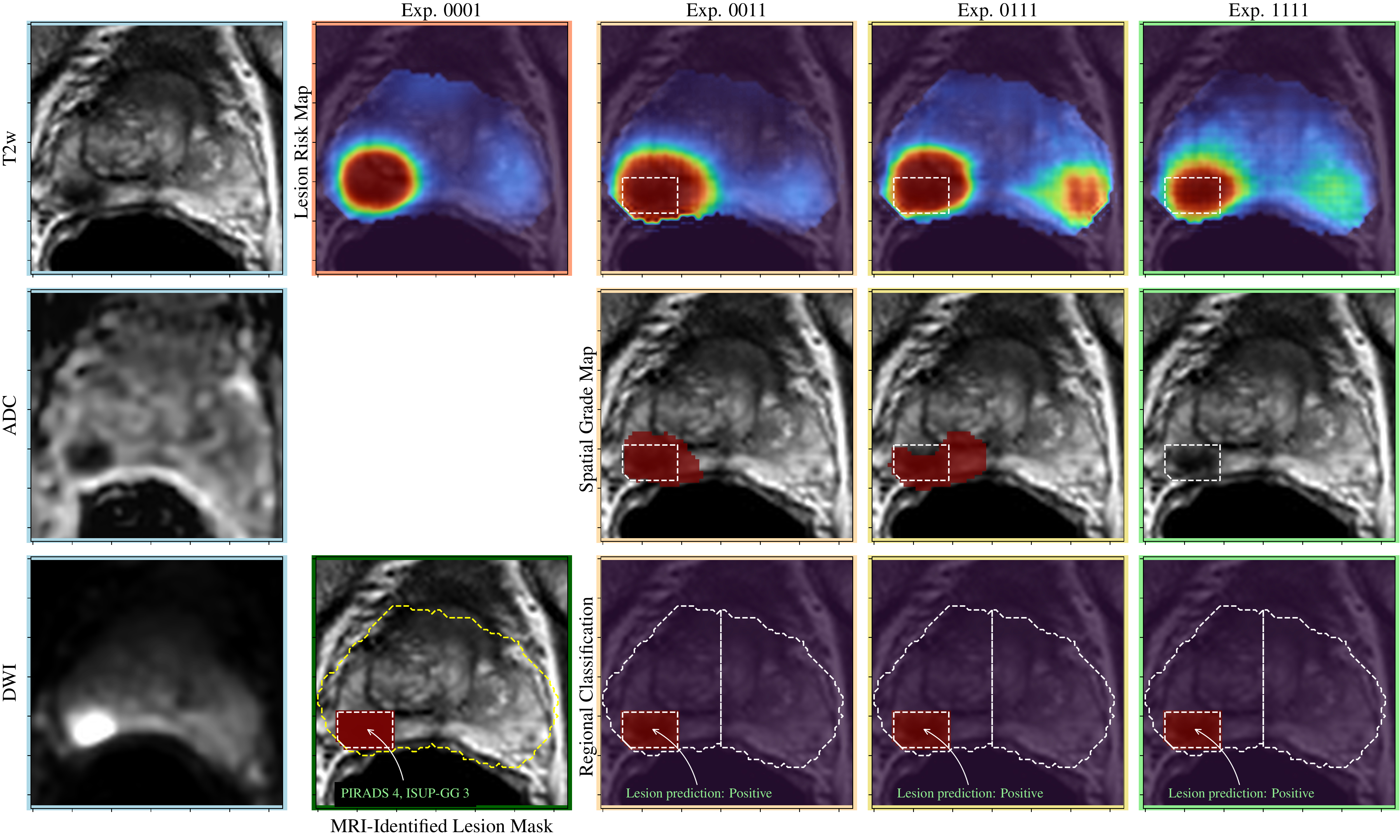}
    \caption{Sample results for a lesion in the right mid posterolateral peripheral zone that was correctly classified as a true positive CS-PCa in all experiments. Note that only a single slice from the 3D volume used is shown. The lesion risk map for all experiments strongly corresponds with the MRI-identified lesion, and the spatial grade maps for experiments 0011 and 0111 also correctly predict CS-PCa. In Experiment 1111, although the visualized slice of spatial grade map does not show CS-PCa, RegionNet achieves a correct classification by picking the highest level of cancer over the entire 3D region.}
    \label{fig:comparison1}
\end{figure*}

\subsection{Comparison of Multi-Task Objectives}
Table~\ref{tab:results} displays the withheld test-set performance of the experiments described in Section~\ref{sec:experiments} with respect to various metrics of success. Experiment~1111, which represents a mixed-supervision objective combining all the aforementioned strong and weak types of supervision, performs the best with respect to lesion classification accuracy as well as gland-wise classification accuracy (based on lesions) compared to the other DNN models, followed by the model trained in Experiment 0111, followed finally by the model in Experiment 0011. This provides strong evidence that weak supervision methods are complementary to strong supervision methods. 

Moreover, the mixed supervision strategies in Experiment~1111 achieve 70.3\% overall (class-balanced) accuracy in classification of clinically significant cancer in lesions, which exceeds the 66.7\% performance of the best PI-RADS cutoff (PI-RADS~$\geq$5) classification assigned by radiologists for this dataset. We believe this is achieved not only by learning from hundreds of datasets, but by incorporation of non-salient cancer histopathology acquired through systematic biopsy.

Table~\ref{tab:results} indicates that although Experiment~1111 performs the best lesion classification, Experiment~0001 achieves the best overlap with radiologists' segmentation of lesions, indicating a trade-off between these two metrics for our dataset.  However, high accuracy of the lesion segmentation is not a major issue in prostate MRI interpretation, as these are relatively readily visualized by radiologists, but the classification of CS-PCa is the more challenging task.
Yet, the radiologist derived lesion segmentation information is still valuable for improving classification performance. For example, Experiment~1110, which includes all mixed-supervision objectives but not lesion segmentation objectives, does not perform as well as Experiment~1111 with respect to lesion classification accuracy (66.2\% vs 70.3\%).  This suggests the lesion segmentation provides some extra attention to suspicious areas during training, improving convergence to desirable minima.
We interpret this lesion segmentation versus classification performance tradeoff in the following way: while lesion segmentation maps and corresponding IoU metrics generally correlate with model performance, they do not dictate the lesion classification performance.

\subsection{Sample Results}
\begin{figure*}[ht]
    \includegraphics[width=\linewidth]{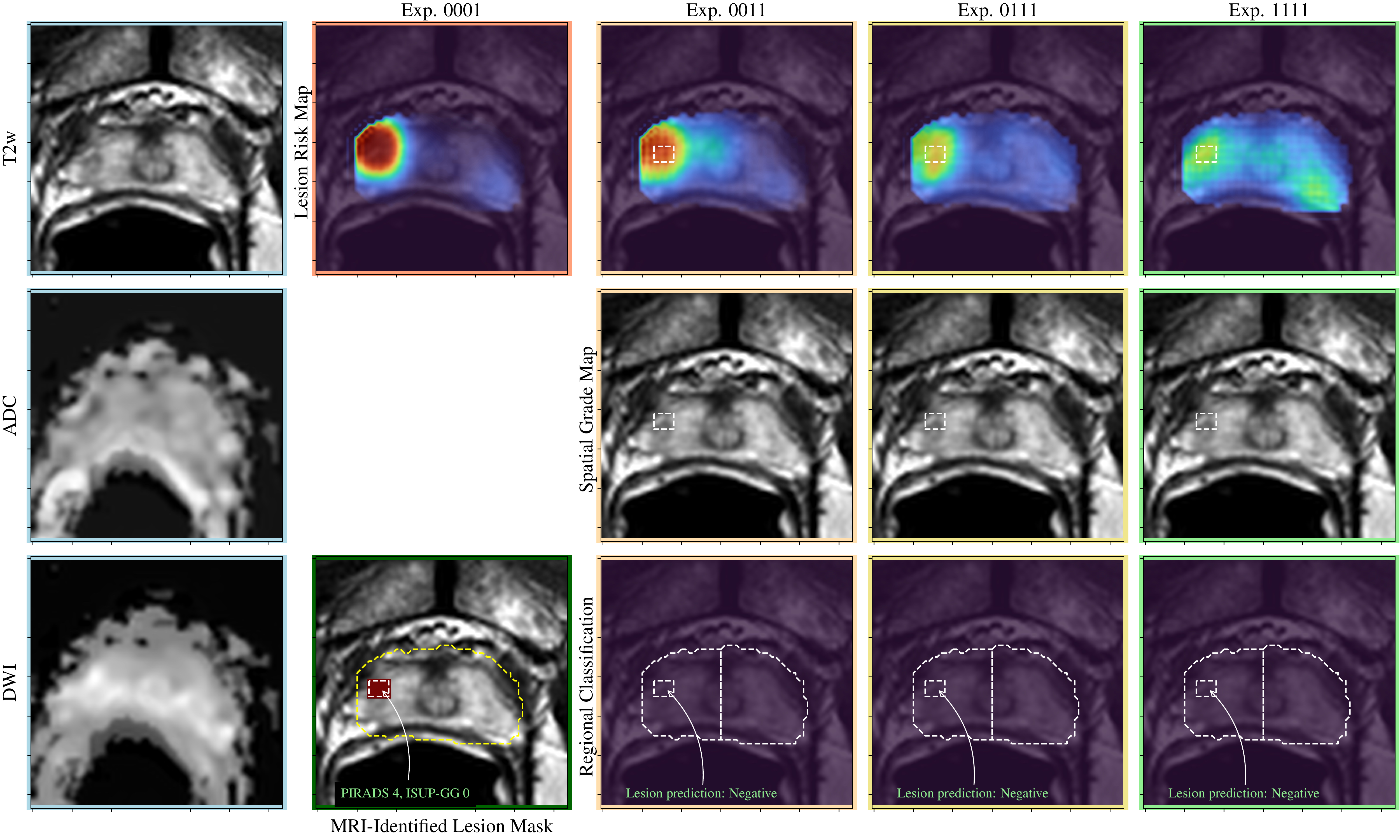}
    \caption{Sample results for a lesion in the right lateral peripheral zone at the apex that was correctly classified as a true negative CS-PCa in all experiments. The lesion risk mask for all for experiments 0001 strongly corresponds to the MRI-identified lesion. This correlation progressively weakens with the subsequent experiments, however the spatial grade map and regional classification correctly predict an absence of CS-PCa.}
    \label{fig:comparison2}
    
    \includegraphics[width=\linewidth]{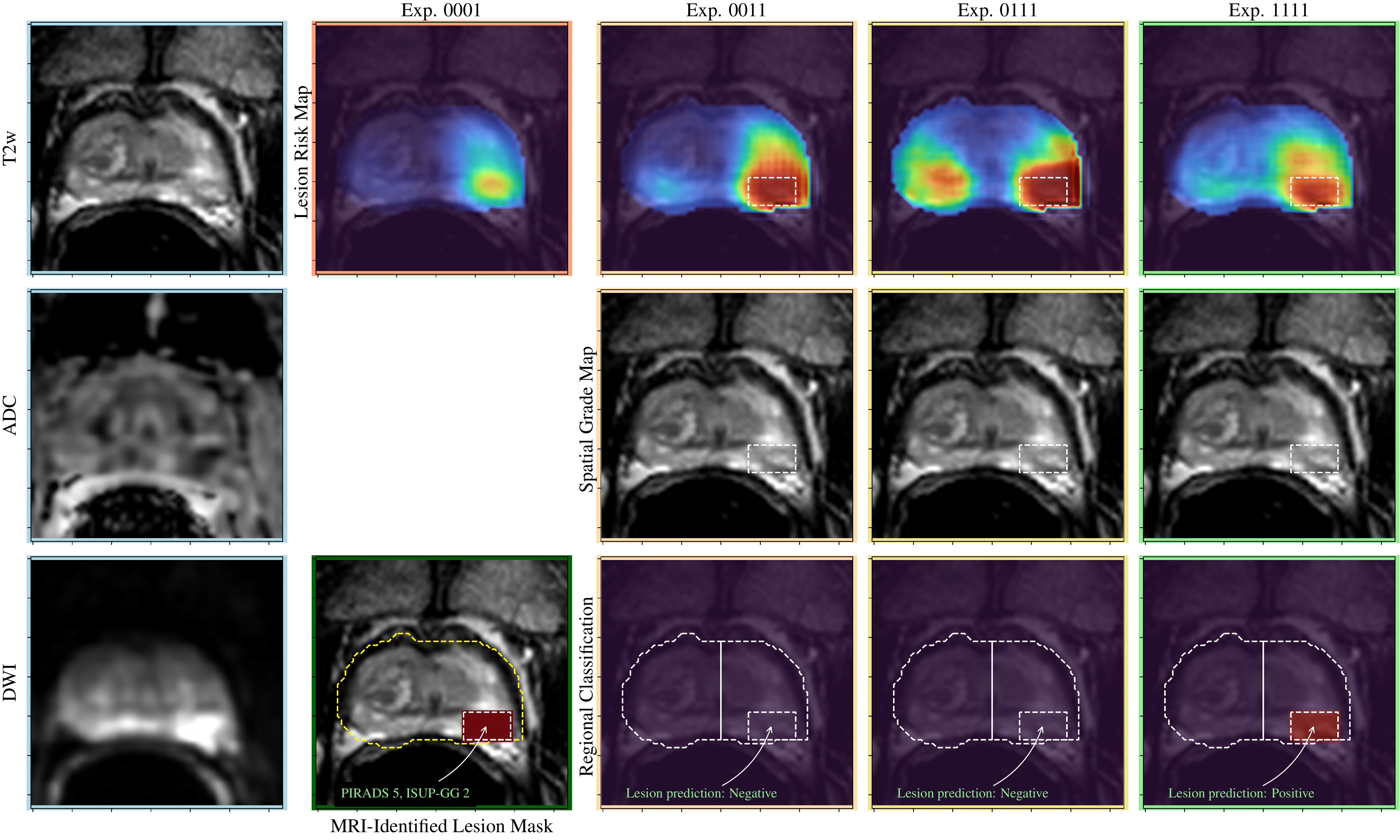}
    \caption{Sample results for a lesion in the left posterolateral peripheral zone that was only correctly classified as a true positive CS-PCa with the addition of weak supervision objectives.  The lesion is identified in all experiments in the Lesion Risk Maps, but was not predicted as CS-PCa until the RegionNet regional classification was added in Experiment 1111.}
    \label{fig:comparison4}
\end{figure*}

A visual comparison of the experiments described in Section~\ref{sec:experiments} is shown in Figures~\ref{fig:comparison1}-\ref{fig:comparison2}, including the pre-processed input images and the various outputs which include a lesion risk map, predicted spatial grading map, and results of the regional classification.

In Figure~\ref{fig:comparison1}, an example is shown where a PI-RADS 4 lesion is identified by the model that is T2 hypointense, markedly hypointense on the ADC map and demonstrating diffusion restriction (high signal intensity) in the right mid posterolateral peripheral zone. This lesion was correctly predicted as clinically significant prostate cancer in all experiments, which corresponds to a biopsy result of ISUP grade group 3.

In Figure~\ref{fig:comparison2}, an example is shown where a PI-RADS-4 lesion is identified in the right lateral peripheral zone at the apex that is T2 hypointense, hypointense on ADC, and demonstrating restricted diffusion (high signal intensity). This lesion is  predicted as not being CS-PCa on all experiments, which corresponds to a biopsy result of benign.

In Figure~\ref{fig:comparison4}, an example is shown where there is a PI-RADS 5 lesion in the left posterolateral peripheral zone that is subtle to detect on T2 weighted imaging, hypointense on the ADC map and demonstrating restricted diffusion (high signal intensity). In experiments 0011 and 0111, the lesion is accurately segmented but not predicted as clinically significant prostate cancer. With the addition of weak supervision methods in experiment 1111, the lesion is predicted to be CS-PCa, corresponding to the biopsy result of an ISUP grade group 2. Thus, this represents a false negative case that was converted to a true positive case with the addition of weak supervision methods.

\section{Discussion}
The results show the benefits of combining strong and weak supervision of pathology into UCNet deep learning model to predict clinically-significant prostate cancer, most clearly summarized in Table~\ref{tab:results}.
This is somewhat intuitive, as we expect improved accuracy by incorporating a much larger set of training labels. However, taking advantage of both heterogeneous pathology ground truth in the form of strongly supervised targeted biopsy and weakly supervised systematic biopsy results required a novel architecture, training, and evaluation methods.
Our results indicate that UCNet was successfully able to take advantage of these diverse sources of training data. This is important, as the inclusion of systematic biopsy procedures continues to be debated clinically~\cite{hugosson2022prostate}.

Our results are particularly interesting because the \textit{lesion-wise} classification accuracy is improved by utilizing ground truth derived from poorly localized systematic biopsy data from tissue that did not appear cancerous on MRI. While this trend seems to occur at the expense of reduced concordance with radiologist-derived MRI target segmentation, lesion classification is the key challenge, not segmentation. Furthermore, 
we note that the current dataset includes 2D bounding boxes for each lesion over the applicable axial T2WI MRI slices which systematically overestimates the lesion size.  The absolute IoU metrics must be put in context then, as they will not be comparable to more finely contoured lesion segmentations, such as those in~\cite{redekop2021attention}.

In our experiments we also observed improved performance by classifying lesions based on the highest grade of cancer rather than the most prevalent cancer. This is shown by comparing Experiment~0111 and Experiment~1111, where the latter had improved performance (improved sensitivity) based on the change to using the RegionNet classifier which applies a bias to threshold the predictions of ISUP grade present along with a most-significant-bit (MSB) classification that picks out the highest grade of cancer predicted to be present. This approach also matches how pathologists interpret biopsy samples, where scoring is done based on the highest cancer grades observed. We believe this classification technique allows the deep learning model to preserve the \textit{distribution} of cancer. This is achieved mechanistically by providing flexibility when interpreting the voxel-based representation, which matched accordingly with the expected uncertainty of the histopathology ground truth. It is also this approach that expands the number of exams and types of ground truth that can be used during training via the mixed supervision objective.

Furthermore, through mixed supervision the UCNet lesion classifier significantly outperforms (70.3\%) the  MRI-based lesion reporting and classification used by radiologists based on PI-RADs (PI-RADS$\geq$5~66.7\%, PI-RADS$\geq$4~66.7\%). Note that although PI-RADs$\geq$5 achieves better performance than PI-RADs$\geq$4, this typically targets only the most obvious MRI-visible lesions, further emphasizing the superior performance of UCNet for classifying less obvious disease. This suggests that there is additional information in prostate MRI images apart from those considered by radiologists using the PI-RADS standardized reporting system that can improve performance. While UCNet's improvement over PI-RADS$\geq$4 is minimal for Gland Accuracy (+1.8\%), we note that UCNet with mixed supervision is able to improve the balance between specificity and sensitivity compared with all other methods (both human and DNN-based), addressing many of the concerns raised by the community~\cite{belue2022low,roest2022comparative}.

This study has limitations. First, this work was conducted with single institutional data, which typically raises concerns over model generalizability.  However, this study uses one of the largest prostate-MRI datasets ($\sim$3x larger than Prostate-X~\cite{litjens2017survey}) with real world clinical data variability including varying acquisition protocols and usage of endorectal coil. Furthermore, our dataset uses the most common approach of combined fusion and systematic biopsy, while Prostate-X used direct MRI-guided biopsy only. Overall, we believe our real world dataset will result in improved generalizability compared to models trained on smaller, more homogeneous datasets. Conscious of this, our approach is designed to enable scaling to diverse multi-institutional datasets with diverse histopathology sources, which can be explored in future work.

Second, in our dataset we had to approximate the systematic biopsy result locations by geometrically splitting the prostate gland into six sextants to represent systematic biopsies, and by using retrospective MRI-based segmentation to represent lesions. This has the potential to cause errors in the assignment of ISUP grade groups to each region as the exact location of systematic biopsy within the approximated region was not known.  If the ultrasound-guidance images were available they could potentially be used to localize the biopsy core by needle position within the gland, as suggested in~\cite{filson2016prostate}. Other datasets, such as Prostate-X, used direct MR-guided biopsy data that potentially provides more accurate localization, although it is also more limited given the absence of systematic biopsy data.

Finally, we emphasize here that MRI-based prostate cancer classification is a difficult task and the major motivating for this project. The current radiology standard of PI-RADS to classify lesions based on size and signal characteristics has relatively poor performance~\cite{westphalen2020variability}. 
Prior work based on radiomics, machine learning, and deep learning have also shown promise but also relatively modest performance improvements~\cite{turkbey2022deep}.
For our dataset, a sophisticated approach was needed to outperform PI-RADS using both targeted fusion biopsy and nontargeted systematic biopsy results. We are optimistic that continued algorithm improvements as well as accumulation of larger, high quality but diverse datasets will further improve the detection of clinically-significant prostate cancer and ultimately impact patient management.

\section{Conclusion}
The proposed mixed supervision methods improved the lesion-wise accuracy of machine learning-based prostate cancer classification algorithms by enabling use of more, diverse histopathology data. We demonstrated this technique by augmenting a popular deep neural network architecture, 3D residual UNet, using regional-query, soft histograms, and a simple threshold-based classification module that enables simultaneous use of voxel- and region-level histopathology ground truth, if available. This is particularly useful for improving classification since systematic biopsy results are typically disregarded in favor of MRI-targeted lesion biopsy results. The results demonstrated a significant improvement in lesion-classification accuracy using mixed supervision objectives, exceeding PI-RADS on a large and diverse dataset of 973 prostate cancer patients.

\section*{Acknowledgements}
This work was supported by NIH/NIBIB grant \#F32EB030411, NIH/NCI grant \#R01CA229354, the UCSF Benioff Initiative for Prostate Cancer Research, a Society for Abdominal Radiology Morton A. Bosniak Research Award, a RSNA Research Resident/Fellow Grant, the Cancer League and Helen Diller Family Comprehensive Cancer Center at UCSF.

\bibliographystyle{ieeetr}
\bibliography{refs}
\end{document}